\documentclass{egpubl}

\ConferencePaper        
 \electronicVersion 

\ifpdf \usepackage[pdftex]{graphicx} \pdfcompresslevel=9
\else \usepackage[dvips]{graphicx} \fi

\PrintedOrElectronic

\usepackage{t1enc,dfadobe}

\usepackage{egweblnk}
\usepackage{cite}
\usepackage{txfonts}
\usepackage{mathrsfs}

\newcommand{\hide}[1]{}
\newcommand{\eg}{\emph{e.g.}, }
\newcommand{\etal}{\emph{et al.}}
\newcommand{\R}{\varmathbb{R}}

\title[Hybrid Lagrangian-Eulerian Model]%
      {A Hybrid Lagrangian-Eulerian Model for the Structural Analysis of Multifield Datasets}

\author[Z. Ding \& X.\,M. Tricoche ]
       {Z. Ding$^{1}$ and X.\,M. Tricoche$^{1}$
        \\
         $^1$Computer Science Department, Purdue University, USA
       }

\begin{document}


\maketitle

\begin{abstract}
Multifields datasets are common in a large number of research and engineering applications of computational science. The effective visualization of the corresponding datasets can facilitate their analysis by elucidating the complex and dynamic interactions that exist between the attributes that describe the physics of the  phenomenon. We present in this paper a new hybrid Lagrangian-Eulerian model that extends existing Lagrangian visualization techniques to the analysis of  multifields problems. In particular, our approach factors in the entire data space to reveal the structure of multifield datasets, thereby combining both Eulerian and Lagrangian perspectives. We evaluate our technique in the context of several fluid dynamics applications. Our results indicate that our proposed approach is able to characterize important structural features that are missed by existing methods.\\


\end{abstract}

\section{Introduction}
\label{sec:introduction}

Multifield datasets, which are comprised of several scalar, vector, or tensor fields, each corresponding to a different physical attribute, are common in a large number of applications of computational science. In particular, physical models based upon partial differential equations naturally produce multifield datasets. For example, in fluid dynamics (CFD) the Navier-Stokes equations combine scalar fields (\eg{} pressure, temperature, density), vector fields (\eg{} velocity, body acceleration), and tensor fields (stress) in the modeling of fluid flows. Similarly, the Maxwell equations state the laws of electromagnetism through a number of scalar (\eg{} charge density), vector (\eg{} electric and magnetic fields), and tensor fields (\eg{} permittivity). 

To enable new insight into multifield datasets, visualization methods must effectively convey the complex and dynamic interactions that exist between their individual attributes. Yet this task is made challenging by the high-dimensionality of the data space spanned by the available fields. 

In this paper, we propose a new approach that combines Eulerian and Lagrangian perspectives and integrate them into a continuum of scales. Specifically, our model analyzes a multifield dataset through the footprint of its spatio-temporal dynamics in the data space. This paper makes following contributions. First, we introduce a new hybrid perspective to analyze the interaction between multiple physical attributes from the viewpoint of dynamic transport in multifield datasets. In addition we present a structure characterization method that derives salient geometric features from this hybrid Lagrangian-Eulerian processing. Finally, we document the benefits of this general framework in the visual analysis of several CFD datasets, whereby our results reveal important structural properties that eluded prior investigations.
%

The remainder of this paper is organized as follows. We first discuss related work in multifield visualization in Section~\ref{sec:related work}. We then describe our proposed Lagrangian-Eulerian approach in Section~\ref{sec:perspective} and comment on implementation details in Section~\ref{sec:implementation}.
We present results for a variety of CFD datasets and offer quantitative comparisons with existing methods in Section~\ref{sec:results}.
Finally, conclusion and future research directions are discussed in Section~\ref{sec:conclusion}.

%

\section{Related Work}
\label{sec:related work}

We discuss in this section the prior work that is most directly related to the present approach, and in particular work that aims to extract features from multifield datasets and to analyze the correlation between individual physical attributes.

To tackle the inherent issue of visual complexity in multifield visualization, data space brushing was proposed for interactive exploration of multifield data~\cite{Martin:1995:High}. Doleisch and Hauser~\cite{Doleisch:2001:Smooth} combined brushing with a transfer function to show interesting flow features. J\"anicke \emph{et al.}~\cite{Janicke:2008:Brushing} proposed a transformation from the high-dimensional data space to a 2D space that preserves proximity. Brushing on the resulting point cloud makes it possible to highlight interesting structures. While brushing allows one to selectively probe different portions of the data space, it does not allow for an overall view of the structural contents of the data and suffers from clutter and occlusion. In addition, brushing implicitly assumes that structures form connected regions in data space and offers no explicit support for the identification of spatially connected features.

Correlation analysis was used to understand and display the relations between fields. Kniss \emph{et al.}~\cite{Kniss:2001:Interactive} presented a method to combine multiple fields through compositing. Sauber \emph{et al.}~\cite{Sauber:2006:Multifield-Graphs:} proposed a gradient similarity measure and a local correlation coefficient to visualize relationships in 3D multifields. Gosink \emph{et al.}~\cite{Gosink:2007:Variable} defined a correlation field as the normalized dot product between two gradient fields from two variables. The derived field was used to study variable interactions with a third variable. Qu \emph{et al.}~\cite{Qu:2007:Visual} introduced the standard correlation coefficient for calculating the correlation strengths between different data attributes in weather data analysis. They presented a weighted complete graph where nodes corresponds to data attributes and weights encode the strength of correlation. While these methods give insight into the \emph{correlation} between different variables, the visualization of the variables themselves remains an open question.

A standard approach to reduce the complexity of large datasets is to focus on their most remarkable features. While a significant literature has been dedicated to the topic in the context of individual fields, only few techniques exist that are suitable for multifield datasets.
J\"anicke \emph{et al.}~\cite{Janicke:2007:Multifield,Janicke:2008:Automatic} introduced methods for multifield data reduction based on statistical complexity. 
A visualization method based on block-wise importance analysis of the data in the joint feature-temporal space was discussed by Wang \emph{et al.}~\cite{Wang:2008:Importance-Driven}. The same authors also analyzed causal relations through information transfer~\cite{Wang:2011:Analyzing}. To identify salient surfaces in multifield datasets, Barakat \etal{}~\cite{Barakat:2012:Surface-Based} applied a fusion approach to merge ridge surfaces extracted from individual fields into a geometric skeleton. They also studied the correlation between different quantities by measuring their joint contribution to individual structures. More closely related to our approach, Shi~\etal{}~\cite{Shi:2008:Visualizing} presented a technique that computes a filtered (\eg{} average) value of a scalar attribute along trajectories to visualize as a color map the transport structure of that quantity. While this method has in common with ours the Lagrangian processing of the data space, it can only handle a single scalar quantity along with the vector field. In contrast our proposed solution is applicable to multiple fields and different data types, its characterization of the distribution of a Eulerian attribute along the flow is not limited to a single scalar measure, and it explicitly characterizes the geometric structure of the multifield.



Finally, query-driven visualization techniques were first discussed by Stockinger \emph{et al.}\cite{Stockinger:2005:DEX:} to reduce the computational complexity of visualization. A subsequent work used correlation fields to explore variable interactions within the domain space of query-regions~\cite{Gosink:2007:Variable}. However, structures of interest may require more complex definition than range queries.

\section{Hybrid Lagrangian-Eulerian Perspective}
\label{sec:perspective}

We present in this section the hybrid Lagrangian-Eulerian perspective that we apply to the characterization of the geometric structure of multifield datasets.

\subsection{Motivation}
\label{ssec:motivatin}

As pointed out in Section~\ref{sec:related work}, most of the methods proposed so far to identify structures in multifield datasets adopt a fundamentally \emph{Eulerian} perspective in their analysis of the data. In other words, they consider the set of physical attributes as anchored in the spatial domain and essentially ignore the spatial dynamics (or transport characteristics) of the phenomenon. 

In contrast, the visualization literature has seen over the last decade so-called Lagrangian methods gain significant prominence, owing to recent advances in fluid dynamics research that have documented their ability to reveal fundamental flow structures such as transport barriers~\cite{Haller:2001:Distinguished}, vortices~\cite{Haller:2016:Defining}, or flow separation~\cite{Weldon:2008:Experimental}. The common aspect of all these methods is the central role played by the flow map, which describes the transport of massless tracers along the flow. 

Motivated by these observations, we propose in this work to reconcile the Eulerian and Lagrangian perspectives and combine them in a hybrid framework in which the transport behavior of the considered multifield dataset affords a Lagrangian lens through which the high-dimensional space of physical attributes can be analyzed. As previously noted, the only existing technique that combines Lagrangian and Eulerian perspectives~\cite{Shi:2008:Visualizing} can only visualize the spatial transport of a single scalar attribute.

\subsection{Definitions and Setup}
\label{ssec:proposed model}

The basic idea of our proposed solution consists in analyzing the footprint of the Lagrangian transport in the high-dimensional data space obtained by compounding the various attributes that jointly form a multifield dataset. Specifically, we consider a multifield dataset $\mathbf{D} = \left(\mathbf{v},\,f_0,\,f_1,\,...,\,f_{n-1}\right)$, where $\mathbf{v}:\,(I \subset \R) \times (D \subset \R^3)  \rightarrow \R^3$ is a smooth three-dimensional time-dependent vector field defined over a spatial domain $D$ and a time interval $I$. The $f_i$ correspond to individual three-dimensional time-dependent scalar, vector, or tensor fields defined over the same spatio-temporal domain $I \times D$ as $\mathbf{v}$. The combined image space of these fields define the high-dimensional data space.

The dynamical system associated with $\mathbf{v}$ describes the motion of massless particles along the flow:
\begin{eqnarray} 
\left\{
\begin{array}{rcl}
\dot{\mathbf{x}}(t,\, t_0,\, \mathbf{x_0}) & = & \mathbf{v}(t,\, \mathbf{x}(t,\, t_0,\, \mathbf{x_0}))\\
\mathbf{x}(t_0,\, t_0,\, \mathbf{x_0}) & = & \mathbf{x_0}.
\end{array}
\right.
\label{eq:dynamical_system}
\end{eqnarray}
The map $\mathbf{x}(\cdot,\, t_0,\, \mathbf{x}_0) : t \mapsto \mathbf{x}(t,\, t_0,\, \mathbf{x_0})$ describes a particle \emph{trajectory}. The map $\mathbf{\phi}(t_0, t) := \mathbf{x}(t, \, t_0, \cdot)$ is called \emph{flow map} and $\mathbf{\phi}(t_0, t)(\mathbf{x_0})$ corresponds to the position $\mathbf{x}(t,\, t_0,\, \mathbf{x_0})$ reached at time $t$ by a particle released at $\mathbf{x_0}$ at time $t_0$.

In this setup, our analysis focuses on the trajectories $\mathbf{y}(\cdot,\, t_0,\, \cdot)$ that are obtained by lifting the trajectories of the dynamical system in Equation~\ref{eq:dynamical_system} into its embedding high-dimensional data space: $\mathbf{y}(t, t_0, \mathbf{y}_0)\,=\, \left(\mathbf{x},\,f_0(\mathbf{x}),\,f_1(\mathbf{x}),\,...,\,f_{n-1}(\mathbf{x}) \right)$, where $\mathbf{x}$ was used as shorthand notation for $\mathbf{x}(t, t_0, \mathbf{x_0})$.

\begin{figure}[ht]
 \centering
 \includegraphics[width=0.95\linewidth]{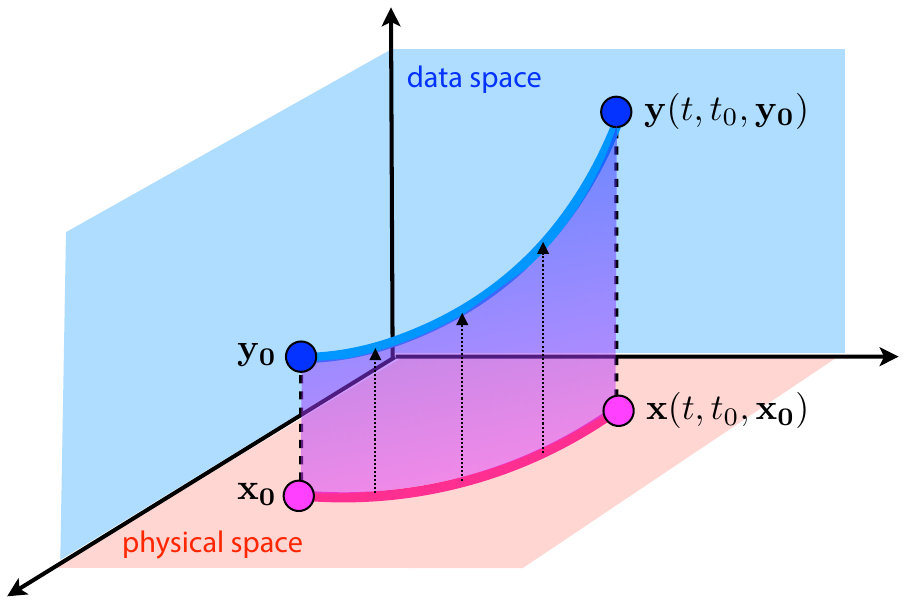}
 \caption{An illustration of Lagrangian-Eulerian concept.}
 \label{fig:lagrangian_eulerian}
\end{figure}

Figure~\ref{fig:lagrangian_eulerian} illustrates the above Lagrangian-Eulerian trajectory definition. The trajectory in physical space is represented by a magenta line. To each point on this trajectory corresponds a point on associated trajectory in the high-dimensional data space, shown in blue.

\subsection{Feature Representation}
\label{ssec:feature}

With the notations above, our approach in this paper consists in performing the structural analysis of the lifted set of trajectories $\mathbf{y}(\cdot, t_0, \cdot)$. Specifically, we consider the mapping $\mathbf{\varphi}$ between the spatio-temporal domain of definition of the multifield and the space of  trajectories in high-dimensional data space: $\mathbf{\varphi}:\,\left(t_0, \mathbf{x_0}\right) \in I \subset \R \times U \subset \R^3 \mapsto \mathbf{y}\left(\cdot, t_0, \mathbf{y_0}\right)$, that is the image of $\mathbf{\varphi}$ is a trajectory in data space seeded at $t_0$ at $\mathbf{y_0}=\left(\mathbf{x_0},\,f_0(\mathbf{x_0}),\,f_1(\mathbf{x_0}),\,...,\,f_{n-1}(\mathbf{x_0}) \right)$.

Indeed, for a given value of $t_0$ we evaluate the mapping $\mathbf{\varphi_d}$ at the vertices of a discretization of the domain $U\subset\R^3$ as the discrete sampling of the corresponding trajectory in a series of successive positions in data space. However, instead of manipulating directly this potentially very large array in subsequent processing, we first transform it into a more convenient, low-dimensional feature representation that we denote $\mathbf{\mu}$ in the following. Practically, we use moments to produce this alternative representation and consider only the first few moments to create the low-dimensional vector values that we associate with each discrete sampling location in the spatial domain. For instance the first two (normalized) moments correspond to the mean and the square root of the covariance matrix of the high-dimensional positions in data space that form a trajectory.

\subsection{Structural Analysis}
\label{ssec:structural analysis}

The final step of our analysis consists in extracting the geometric structure of the vector-valued representation of the multifield dataset that we created. We define this structure as a skeleton comprised of \emph{edges}, in other words a set of surfaces along which the vector-valued dataset exhibits discontinuities.

Adapting to our setting the general strategy developed in the image processing literature to characterize edges in images, we evaluate at each location the magnitude of the gradient (spatial derivative) of the vector-valued dataset. To ensure the robustness of our computation, we evaluate this gradient at each location through a least squares linear fit of the vector-valued mapping in its 1-neighborhood. Here the function to minimize can be written:
$\sum_{k \in \mathscr{N}_1(i)} || \mathbf{\mu_k} - \mathbf{\mu_i} - A\,(\mathbf{x_k}-\mathbf{x_i})||^2$, where $\mathscr{N}(1)(i)$ is the set of 1-neighbors of a given vertex $i$, $\mathbf{\mu_k}$ is the feature vector value at a neighbor $\mathbf{x_k}$ and $A$ is a matrix to fit that corresponds to the gradient that we want to compute at $\mathbf{x_i}$. Once this gradient is computed, we measure its norm as the spectral norm of the matrix $A$, which we obtain as its largest singular value. 

\section{Implementation}
\label{sec:implementation}
In this section, we discuss the implementation details of the presented hybrid Lagrangian-Eulerian model.

Without loss of generality, we limit our description here to a two-dimensional uniformly spaced setting. 
Similar constructions are possible in any spatial dimension and on most types of discrete domains.

Assume a point $\mathbf{x}_{i,j}:=(i\cdot h, j\cdot h)$ with $h > 0$ and $i,j\in \left \{ 0,\cdots,N \right \}$ in our discrete setting. 
Furthermore, let $M > 0$ an integer defining a time discretization $t^{m}:=\frac{m/M}{T}$ with $m\in\left \{ 0, \cdots, M \right \}$.
For a pathline $P_{t_{0},\mathbf{x_{i,j}}}^{T}$ generated by tracing the particle on $\mathbf{x}_{i,j}$ we define the discrete pathline point
\begin{equation}
\mathbf{x}^{m} := P_{t_{0},\mathbf{x}_{i,j}}^{T}(t^{m})
\end{equation}
as the discrete representation of the pathline using a numerical integration.

Next, combing the spatial coordinate $\mathbf{x}^{m}$ with time $t_{0} + t^{m}$ we have the discrete trajectory $\mathscr{T}_{t_{0},\mathbf{x}_{i,j}}^{T}$ in space-time domain.
Scalar attribute values associated to every discrete space-time point $(\mathbf{x}^{m}, t_{0} + t^{m})$ are computed by interpolating the input scalar field and form the discrete trajectory $\mathscr{S}_{t_{0},\mathbf{x}_{i,j}}^{T}$ in data space.

In our implementation, the measurement of edge strength in the resulting space filling data space trajectory field is achieved by estimating a function of neighboring trajectories.

For any location $\mathbf{x}$ in the domain of a given dataset, we define our trajectory function by the following equation,
\begin{equation}
\mathscr{S} = A \cdot \mathbf{x} + \mathbf{b}
 \label{eq:trajectory_function}
\end{equation}

Thus, the spatial variations around $\mathbf{x}$ are determined by its Jacobian $J(\mathbf{x}):=\nabla_{\mathbf{x}} \mathscr{S} \equiv A$ and the edge strength is measured as the spectral norm of $J(\mathbf{x})$: 
\begin{equation}
edge(\mathbf{x}):=\sqrt{\lambda_{max}(J(\mathbf{x})^{T}J(\mathbf{x}))}.
\label{eq:le_edge_strength}
\end{equation}

Instead of directly estimating the trajectory function of equation~\ref{eq:trajectory_function},
we map each data space trajectory to a fixed number of parameters and low-dimensional Euclidean feature space. 
A set of moments is employed to capture the statistical properties of a trajectory,
\begin{equation}
S_{t_{0},\mathbf{x}_{i,j}}^{T} = (\mu_{1}, \mu_{2}, \cdots)^\intercal
\end{equation}
where $S_{t_{0},\mathbf{x}_{i,j}}^{T}$ is the parameterization of $\mathscr{S}_{t_{0},\mathbf{x}_{i,j}}^{T}$. 
Equation~\ref{eq:trajectory_function} is then turned into the following equation,
\begin{equation}
S_{t_{0},\mathbf{x}_{i,j}}^{T} = \widetilde{A} \cdot \mathbf{x}_{i,j} + \widetilde{b}.
\label{eq:trajectory_function_param}
\end{equation}

Thus, the edge strength on $\mathbf{x}_{i,j}$ is approximated as
\begin{equation}
edge(\mathbf{x}_{i,j}) \approx \widetilde{edge}(\mathbf{x}_{i,j}) = \sqrt{\lambda_{max}(\widetilde{A}^\intercal \widetilde{A})}.
\end{equation}

\begin{figure}[htbp]
 \centering
 \includegraphics[width=\linewidth]{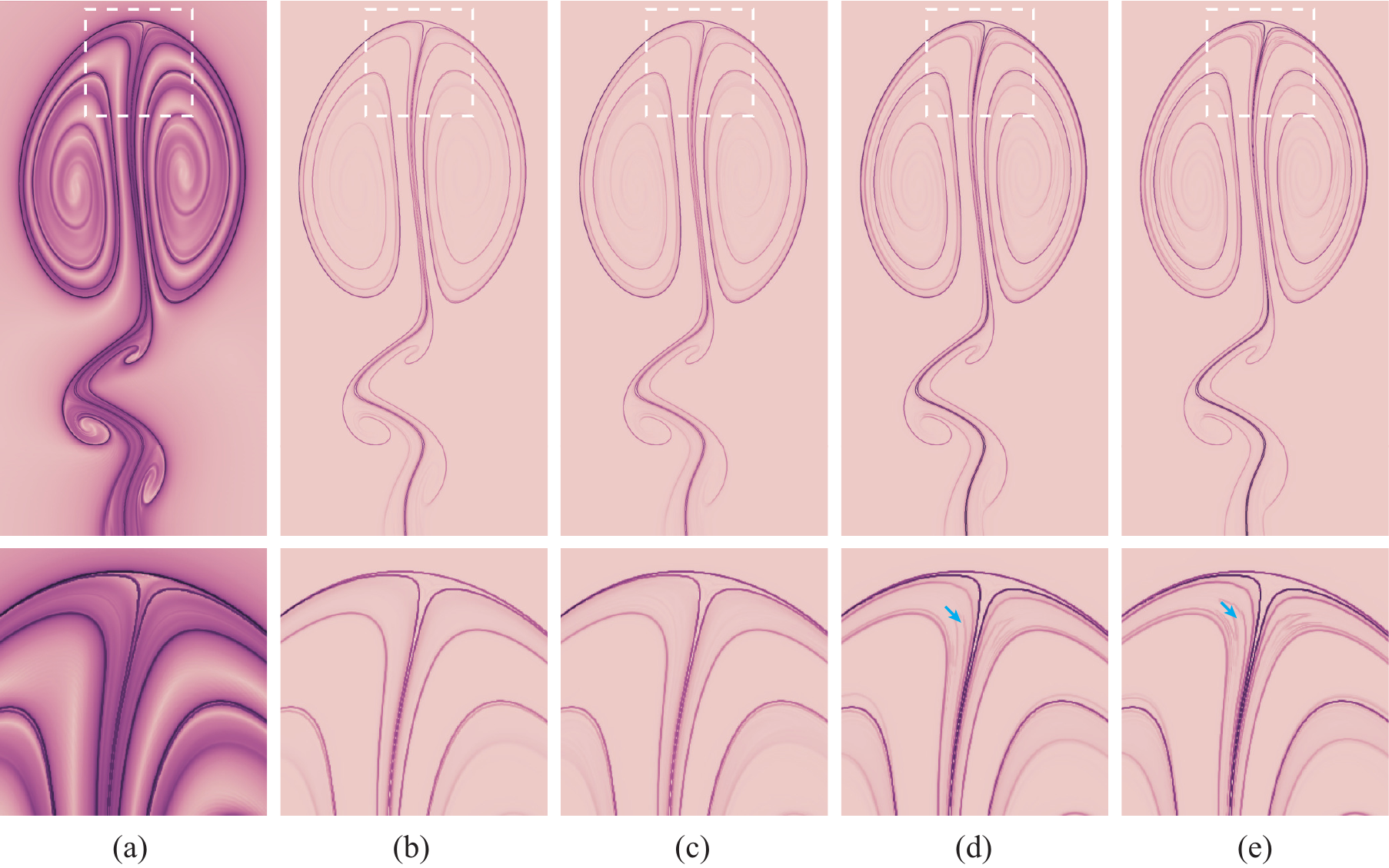}
 \caption[A comparison of edge strength measured using the presented hybrid Lagrangian Eulerian method with different moments during the trajectory parameterization.]{A comparison of edge strength measured using the presented hybrid Lagrangian-Eulerian method with different moments during the trajectory parameterization. (a) Backward FTLE computed from \textit{Boussinesq Flow} dataset with $t_{0}=3.0$ and $T=2.0$; (b)-(e) edge strength measured by the presented approach using up to $1st$, $2nd$, $3rd$, and $5th$ moments during the trajectory parameterization.}
 \label{fig:parameters}
\end{figure}

Figure~\ref{fig:parameters} compares a backward FTLE field computed from \textit{Boussinesq Flow} (Section~\ref{res:boussinesq_flow}) using $t_{0}=3.0$ and $T=2.0$ with the edge strength fields measured by the presented approach with different sets of moments during the trajectory parameterization.
A set with higher order moments not only is hard to estimate but also presents false structures (highlighted by cyan arrows). In our system, we use the first moment (mean) and the second central moment (variance) to map trajectories to the feature space. 

We recorded the running time of our hybrid Lagrangian-Eulerian approach and finite-time Lyapunov exponent (FTLE)~\cite{ZD:Haller2001} method on a machine with an Intel i7-970 processor (6 cores / 12 threads) and 12GB memory to study the performance of both methods.

\begin{table*}[ht]
\centering
\caption{Running time of different datasets.}
\label{tab:running_time}
\begin{tabular}{|c|c|c|c|}
\hline
dataset & \begin{tabular}[c]{@{}c@{}}spatial \\ resolution\end{tabular} & FTLE & Lagrangian-Eulerian \\ \hline
Boussinesq & $800\times 2400$ & 88 minutes & 90 minutes \\ \hline
Hurricane \textit{Isabel} & $500\times 500\times 100$ & 18 minutes & 18 minutes \\ \hline
Delta Wing (right bubble) & $600\times 350\times 400$ & 69 hours & 70 hours \\ \hline
\end{tabular}
\end{table*}

Table~\ref{tab:running_time} summarizes the running time of both methods on different datasets.
The running time of FTLE method and hybrid Lagrangian-Eulerian approach are close to each other. 
It is because the computational time spends on pathline / flowmap generation dominates the running time in both cases.

\section{Results}
\label{sec:results}

We evaluated our method with a number of multi-field datasets from analytic and real-world CFD problems.

\subsection{Boussinesq Flow}
The Boussinesq flow used here was simulated using the example provided in Gerris Flow Solver~\cite{ZD:Popinet2004} by the author. 
The simulation uses the classical Boussinesq approximation, in which the conservation of mass is replaced by the conservation of volume (zero divergence), to generate the turbulent vortex behavior in the physical domain $Space \times Time = [-0.5, 0.5] \times [-0.5, 2.5] \times [0.0, 10.0]$. 
With this approximation, fluid heated by some objects, e.g. the heated cylinder in this case, reduces the mass. 
Thus, it moves upward by the buoyancy effect and creates the turbulence~\cite{ZD:Wilcox1998}. 
This simulation result contains a 2D unsteady velocity field as well as an associated temperature field of every time frame.

\label{res:boussinesq_flow}
\begin{figure}[!htbp]
 \centering
 \includegraphics[width=\linewidth]{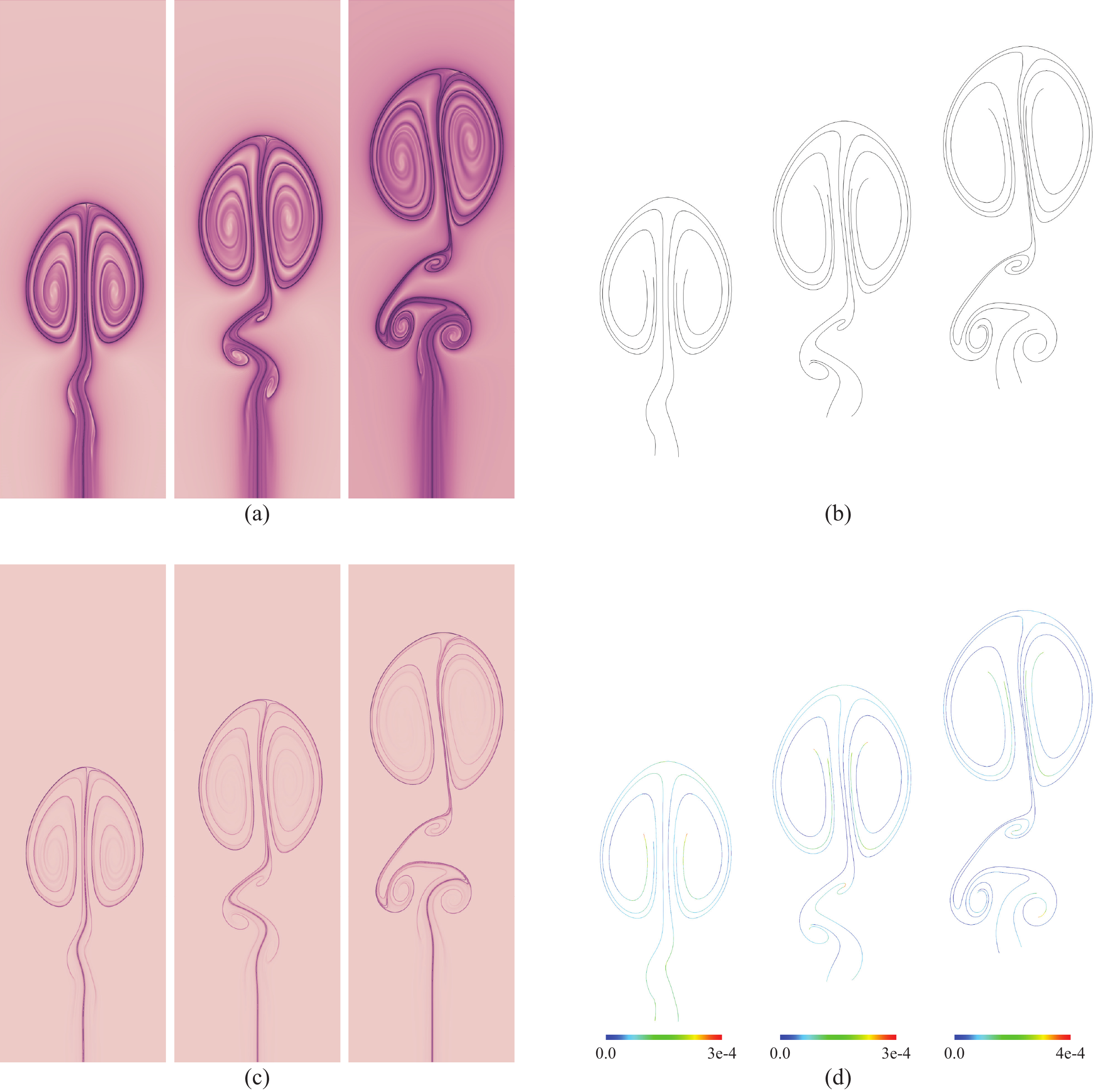}
 \caption[Salient structures characterized in \textit{Boussinesq flow} dataset using the hybrid Lagrangian-Eulerian method.]{Salient structures characterized in \textit{Boussinesq flow} dataset using the hybrid Lagrangian-Eulerian method. (a) three backward FTLE fields computed with different starting time $t_{0}=2.75$, $t_{0}=3.00$, and $t_{0}=3.25$; (b) ridge lines extracted from those FTLE fields; (c) three edge strength fields measured by the presented approach with a same setting as (a); (d) ridge lines extracted from edge strength fields, color coded by the shortest Euclidean distance to FTLE ridges.}
 \label{fig:boussinesq_flow}
\end{figure}

We applied both the FTLE method and the hybrid Lagrangian-Eulerian approach on this dataset and compared them through the similarity of ridge lines extracted from the resulting scalar fields.
Given two ridge lines $C_{i}={\mathbf{c}_{i}^{1}, \cdots, \mathbf{c}_{i}^{m}}$ and $C_{j}={\mathbf{c}_{j}^{1}, \cdots, \mathbf{c}_{j}^{n}}$ that are represented by polylines with $m$ and $n$ vertices respectively, we define the dissimilarity metric $d_{ij}$ as the distance from $C_{i}$ to $C_{j}$,
\begin{equation}
d_{ij}=\frac{1}{m+n}\left [ \sum_{k=1}^{m}dist(\mathbf{c}_{i}^{k}, C_{j})+\sum_{k=1}^{n}dist(\mathbf{c}_{j}^{k}, C_{i}) \right ]
\end{equation}
where the function $dist(\mathbf{c}_{i}^{k},C_{j})$ returns the shortest Euclidean distance from the vertex $\mathbf{c}_{i}^{k}$ to the piecewise linear curve $C_{j}$.
Instead of finding the maximal value in the union set of $\left \{ dist(\mathbf{c}_{i}^{k}, C_{j}):k\in\left \{ 1,\cdots,m \right \} \right \}$ and $\left \{ dist(\mathbf{c}_{j}^{k}, C_{i}):k\in\left \{ 1,\cdots,n \right \} \right \}$ used in Hausdorff distance metric~\cite{ZD:Huttenlocher1993}, our metric uses the mean value of the same union set.

Figure~\ref{fig:boussinesq_flow} (a) illustrates three backward FTLE fields computed with different starting time $t_{0}=2.75$, $t_{0}=3.00$, and $t_{0}=3.25$. 
Figure~\ref{fig:boussinesq_flow} (b) shows ridge lines extracted from those FTLE fields using Marching Ridge method. 
In figure~\ref{fig:boussinesq_flow} (c), three edge strength fields were measured with the same setting as figure~\ref{fig:boussinesq_flow} (a) by the presented hybrid Lagrangian-Eulerian approach. 
We colored each vertex of ridge lines extracted from our edge strength measurement based on the shortest Euclidean distance to FTLE ridges in figure~\ref{fig:boussinesq_flow} (d).
The distances from edge strength ridges to FTLE ridges of the above three results are $7\mathrm{e}{-5}$, $5\mathrm{e}{-5}$, and $6\mathrm{e}{-5}$ and all of them are far less than the distance between neighboring points in our discrete setting ($h=5\mathrm{e}{-4}$).
The comparison on the Boussinesq flow dataset shows in some simple cases, structures characterized by our hybrid Lagrangian-Eulerian approach coincide with the LCS characterized from an FTLE measure.

\subsection{Hurricane Isabel}
The \textit{Hurricane Isabel} is the benchmark dataset of IEEE 2004 Visualization Design Contest. 
This dataset was generated by the National Center for Atmospheric Research in the United States to simulate the \textit{hurricane Isabel} in a large region of the West Atlantic in September 2003.
It consists of $13$ variables including a vector field which is the wind speed and $12$ scalar properties such as temperature, pressure, and precipitation for $48$ time frames ($1$ simulated hour between time steps).
The spatial dimension of the dataset is $500\times 500\times 100$ at each time frame.

The hurricane eye is the most recognizable feature and is considered as the focus of a hurricane.
One major goal was to study the path and the structure of the hurricane eye.
A hurricane eye is the center point about which the rest of the storm rotates and where the lowest pressure is found in the hurricane.
The wind will increase its magnitude and converge towards the hurricane eye.
Before reaching the eye, the \textit{Coriolis force} deflects the wind slightly away from it and quickly drops the wind magnitude, causing the wind to rotate around the center of the hurricane, which is named as the hurricane eye wall.
Thus, tracking and analyzing the hurricane eye wall provides an effective way to study the path and the structure of the hurricane eye. 

\begin{figure}[!htbp]
 \centering
 \includegraphics[width=\linewidth]{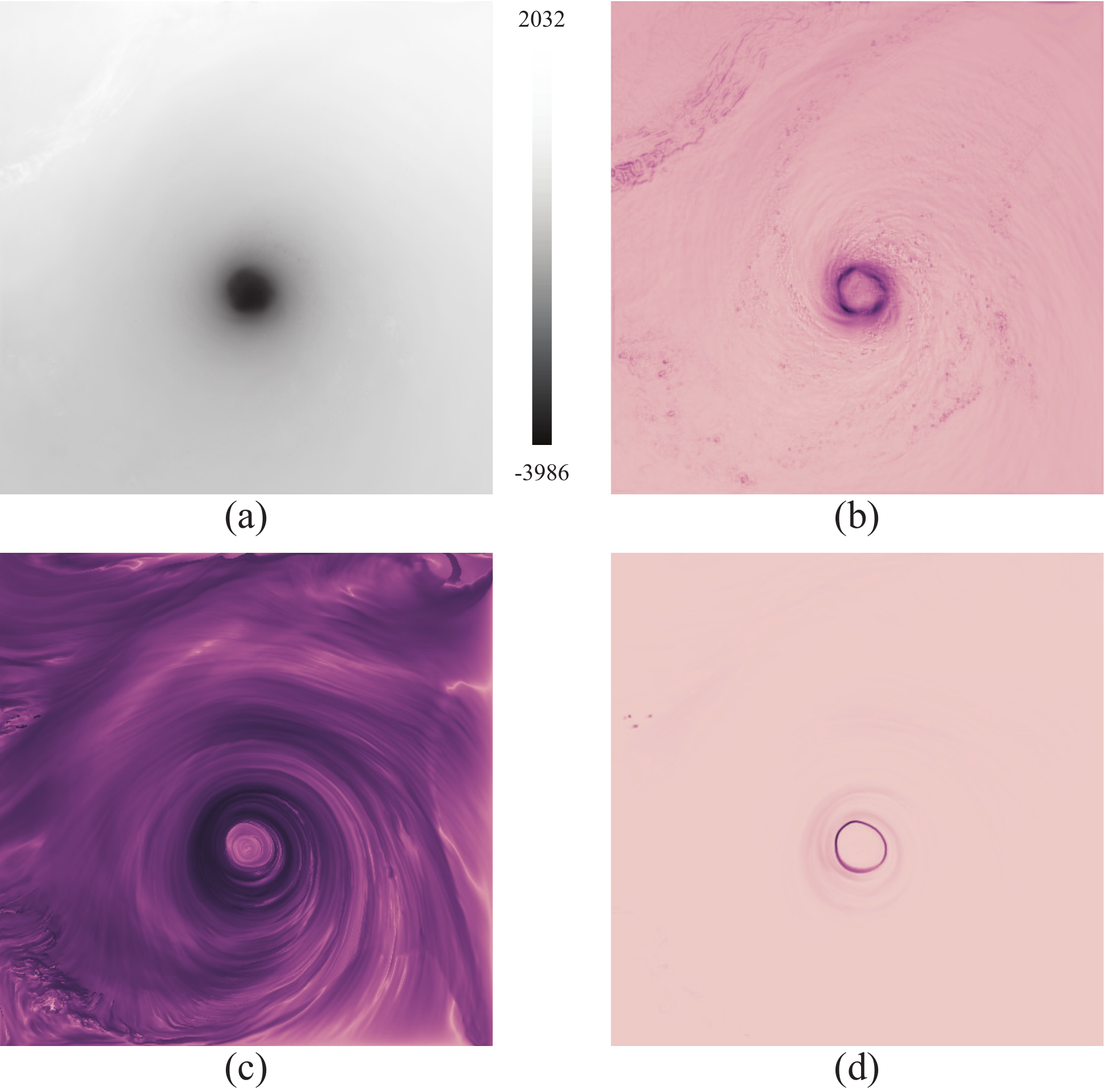}
 \caption[Salient structures characterized in \textit{Hurricane Isabel} dataset using the hybrid Lagrangian-Eulerian method.]{Salient structures characterized in \textit{Hurricane Isabel} dataset using the hybrid Lagrangian-Eulerian method. (a) a slice of the pressure volume; (b) Canny edge detection result; (c) FTLE result; (d) hybrid Lagrangian-Eulerian approach result.}
 \label{fig:hurricane_isabel_slice}
\end{figure}

Figure~\ref{fig:hurricane_isabel_slice} (a) shows a slice of the pressure volume. 
The low pressure region (the black spot) on this image indicates the hurricane eye. 
One way to characterize the hurricane eye wall is applying an edge detection method, e.g. Canny edge detector~\cite{ZD:Canny1986}, on the pressure volume as shown in figure~\ref{fig:hurricane_isabel_slice} (b).
However, the noise near the hurricane eye in the pressure volume results in low quality edges.
Because of the convergence and divergence behavior of the wind around the hurricane eye, the Lagrangian method is another way to capture the geometric structure of the hurricane eye wall. 
However, the nearby rainbands and other structures of the hurricane make it difficult to distinguish the hurricane eye wall from other structures detected inside this dataset.
Figure~\ref{fig:hurricane_isabel_slice} (c) illustrates the same slice of the forward FTLE volume computed with integration time $T=2$ hours as figure~\ref{fig:hurricane_isabel_slice} (a).
Using the pressure as the scalar field, and with a same integration time as the FTLE method, our hybrid Lagrangian-Eulerian approach is able to generate a clear and smooth hurricane eye wall result (figure~\ref{fig:hurricane_isabel_slice} (d)).

\begin{figure}[!htbp]
 \centering
 \includegraphics[width=\linewidth]{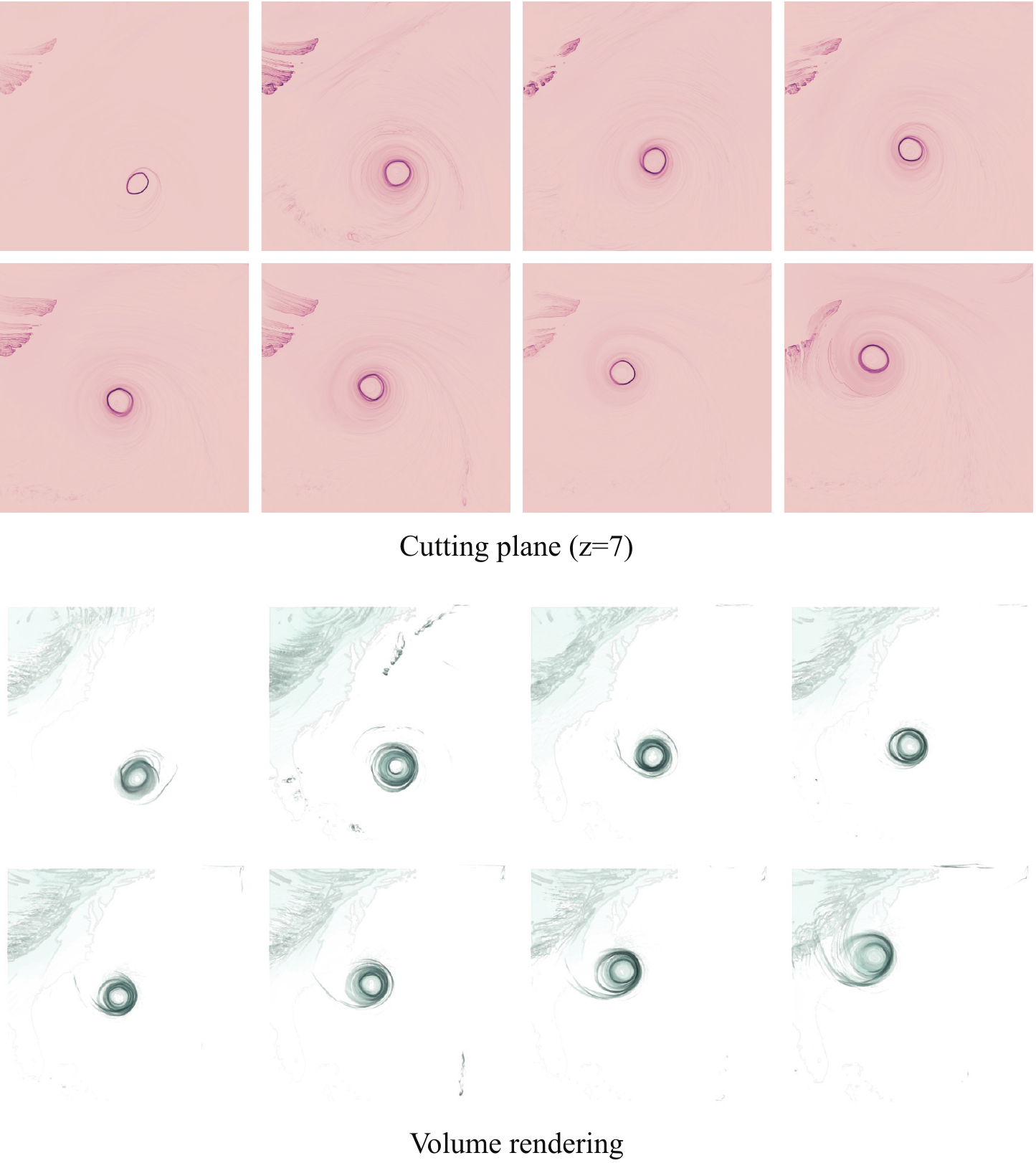}
 \caption[Tracking the hurricane eye wall in \textit{Hurricane Isabel} dataset.]{Tracking the hurricane eye wall in \textit{Hurricane Isabel} dataset.}
 \label{fig:hurricane_isabel_eye_track}
\end{figure}

Figure~\ref{fig:hurricane_isabel_eye_track} visualizes the edge strength measured by the proposed approach using cutting plane and volume rendering. It is clear that with our approach, one can easily track and analyze the hurricane eye wall in this dataset. 

\begin{figure}[!htbp]
 \centering
 \includegraphics[width=\linewidth]{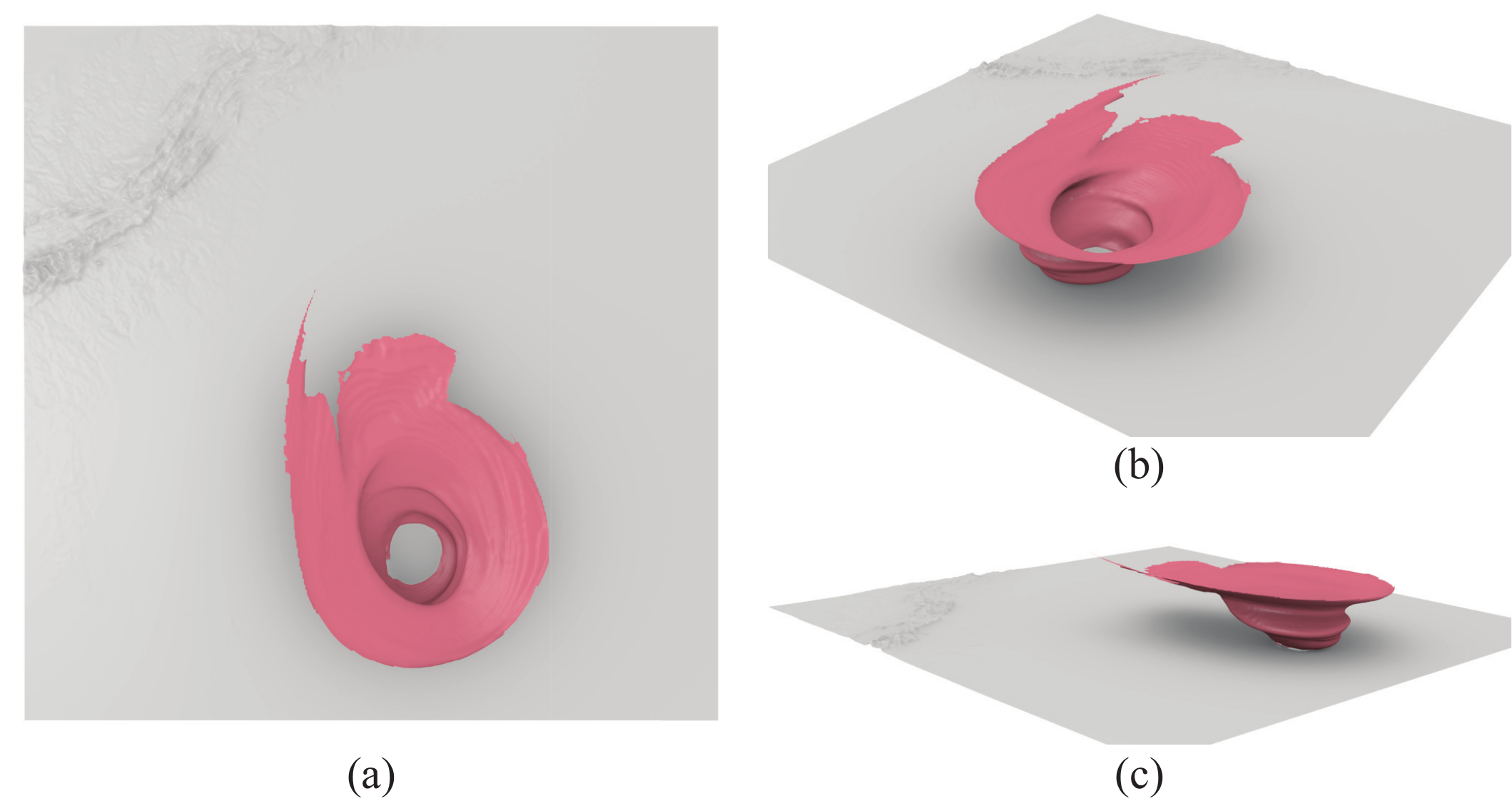}
 \caption[\textit{Hurricane Isabel} results.]{\textit{Hurricane Isabel} results.}
 \label{fig:hurricane_isabel_result}
\end{figure}

Figure~\ref{fig:hurricane_isabel_result} shows the ridge surfaces extracted from the edge strength volume measured by our hybrid Lagrangian-Eulerian approach.
The resulting ridge surfaces successfully capture the structure of the hurricane eye wall in this dataset.

\subsection{Delta Wing}
The so-called Delta Wing dataset is a CFD simulation that is designed to study the transient flow above
a delta wing at low speeds and increasing angle of attack.
The major goal of the simulation was to investigate the cause of the vortex breakdown of the primary vortices.
One velocity field and four scalar attribute fields are included in this dataset.

\begin{figure}[!htbp]
 \centering
 \includegraphics[width=\linewidth]{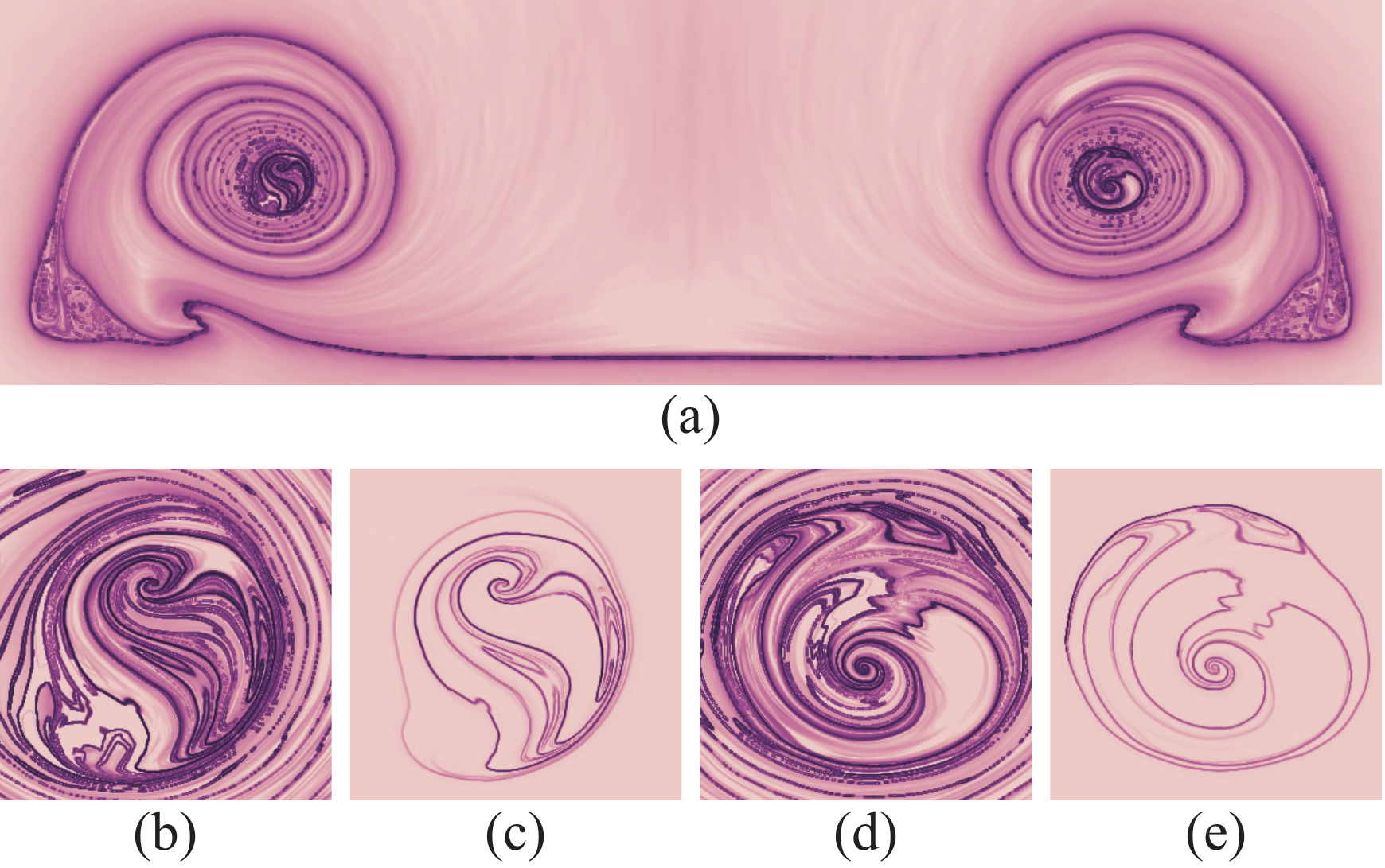}
 \caption[A comparison of FTLE and hybrid Lagrangian-Eulerian method in \textit{Delta wing} dataset.]{A comparison of FTLE and hybrid Lagrangian-Eulerian method in \textit{Delta wing} dataset. (a) One slice of the FTLE field computed from this dataset; (b) and (d) closed-up of FTLE field in both vortex breakdown bubbles; (c) and (e) edge strength measured by the presented hybrid Lagrangian-Euleiran approach of both vortex breakdown bubbles.}
 \label{fig:delta_wing_ftle_le}
\end{figure}

Two asymmetric vortex breakdown bubbles, a chaotic one on the left side and a stable one the right side, have already been visualized using stream surfaces in previous studies~\cite{ZD:Tricoche2004,ZD:Garth2005}.
In the Lagrangian method, e.g. FTLE, the main challenge in characterizing LCS of the vortex breakdown bubble is a noisy FTLE field caused by the complexity of the flow near the edge of the delta wing.
Figure~\ref{fig:delta_wing_ftle_le} (a) shows a slice of the backward FTLE field computed from this dataset.
The detailed FTLE fields in regions of both vortex breakdown bubbles of two primary vortices are shown in figure~\ref{fig:delta_wing_ftle_le} (b) and (d).
The unexpected structures at those regions, especially the ones near the outer layer of both vortex breakdown bubbles, make it difficult to extract smooth and meaningful LCS.

\begin{figure}[!htbp]
 \centering
 \includegraphics[width=\linewidth]{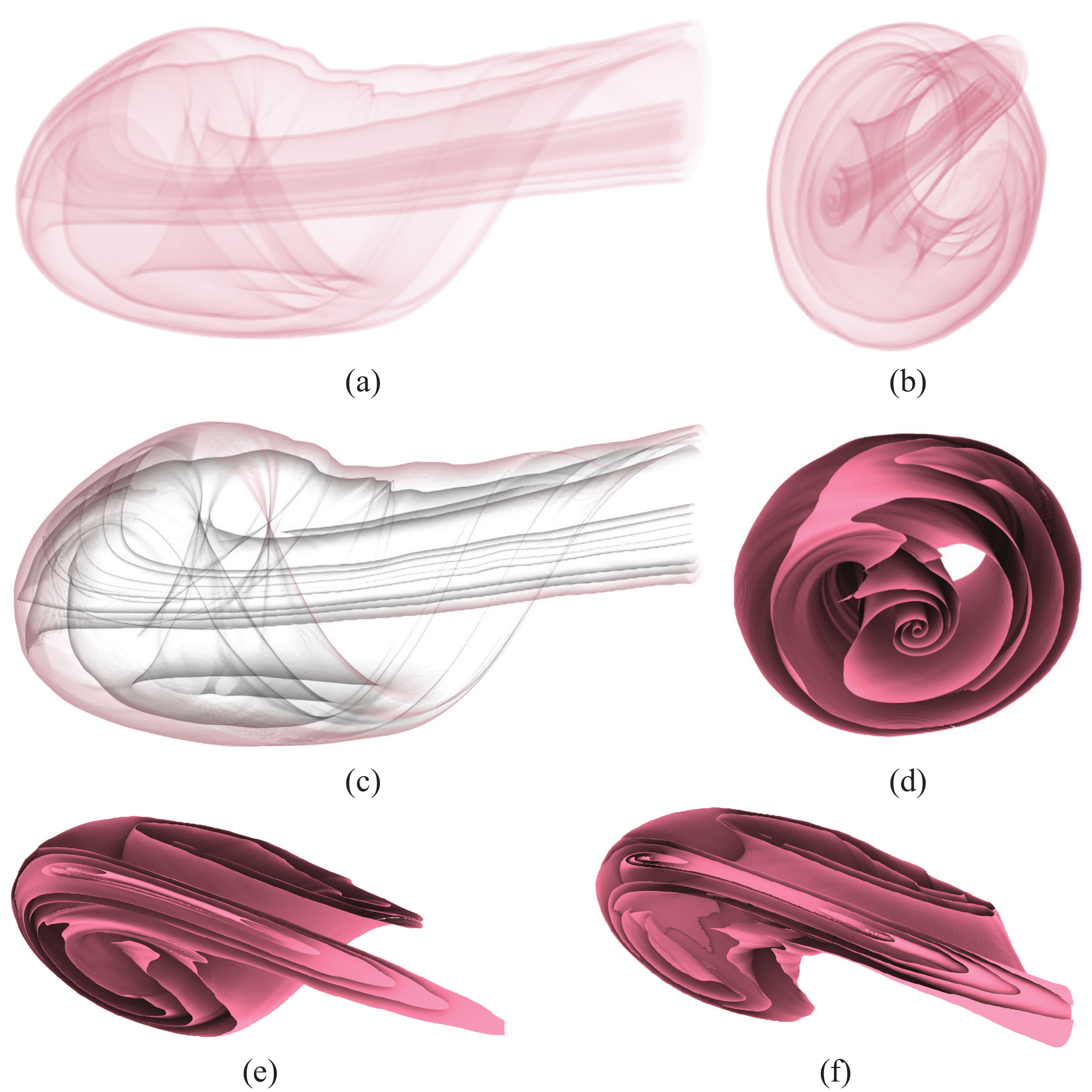}
 \caption[The right vortex breakdown bubble result.]{The right vortex breakdown bubble result. (a) and (b) volume rendering of the edge strength volume; (c)-(f) ridge surfaces characterized from the same edge strength volume, rendered with opacity and different cutting planes.}
 \label{fig:delta_wing_right_bubble}
\end{figure}

Using the SA viscosity, which is computed from Spalart-Allmaras model~\cite{ZD:Spalart1992}, as the associated scalar field in the presented hybrid Lagrangian-Eulerian approach, our edge strength measurement generates a clear result and captures coherent structures of both vortex breakdown bubbles (figure~\ref{fig:delta_wing_ftle_le} (c) and (e)). 
The resulting edge strength volume of the right vortex breakdown bubble are illustrated in figure~\ref{fig:delta_wing_right_bubble} using volume rendering and ridge surfaces.


\section{Conclusion}
\label{sec:conclusion}

In many applications of flow visualization, the characterization of boundaries of coherent regions is a crucial step. In this paper, we have shown that combing generalized notions of Eulerian analysis and Lagrangian analysis enables better structure characterization results. We have evaluated the proposed hybrid Lagrangian-Eulerian model with real-world datasets and documented its ability to capture important features in regions where existing approaches fail. 

An interesting avenue for future work concerns the improvement of the performance of this approach. Both a GPU-based implementation and a pathline reuse strategy would reduce the computational time of our approach.

\bibliographystyle{eg-alpha-doi}

\bibliography{Ziang,xmt}

\newcommand{\etalchar}[1]{$^{#1}$}
\begin{thebibliography}{\uppercase{KMM{\etalchar{*}}01}}

\bibitem[BRT12]{Barakat:2012:Surface-Based}
\textsc{Barakat S.~S., R{\"u}tten M., Tricoche X.}:
\newblock Surface-based structure analysis and visualization for multifield
  time-varying datasets.
\newblock \emph{IEEE Transactions on Visualization and Computer Graphics 18},
  12 (2012), 2392--2401.

\bibitem[Can86]{ZD:Canny1986}
\textsc{Canny J.}:
\newblock A computational approach to edge detection.
\newblock \emph{IEEE Transactions on Pattern Analysis and Machine
  Intelligence}, 6 (1986), 679--698.

\bibitem[DH01]{Doleisch:2001:Smooth}
\textsc{Doleisch H., Hauser H.}:
\newblock Smooth brushing for focus+context visualization of simulation data in
  3d.
\newblock In \emph{Journal of WSCG} (2001), pp.~147--154.

\bibitem[GABJ07]{Gosink:2007:Variable}
\textsc{Gosink L., Anderson J., Bethel W., Joy K.}:
\newblock Variable interactions in query-driven visualization.
\newblock \emph{IEEE Transactions on Visualization and Computer Graphics 13}
  (November 2007), 1400--1407.
\newblock URL: \url{http://dx.doi.org/10.1109/TVCG.2007.70519}, \href
  {http://dx.doi.org/http://dx.doi.org/10.1109/TVCG.2007.70519}
  {\path{doi:http://dx.doi.org/10.1109/TVCG.2007.70519}}.

\bibitem[GT05]{ZD:Garth2005}
\textsc{Garth C., Tricoche X.}:
\newblock Topology- and feature-based flow visualization: Methods and
  applications.
\newblock In \emph{SIAM Conference on Geometric Design and Computing} (2005),
  vol.~17, Citeseer.

\bibitem[Hal01a]{Haller:2001:Distinguished}
\textsc{Haller G.}:
\newblock Distinguished material surfaces and coherent structures in
  three-dimensional flows.
\newblock \emph{Physica D 149} (2001), 248--277.

\bibitem[Hal01b]{ZD:Haller2001}
\textsc{Haller G.}:
\newblock Distinguished material surfaces and coherent structures in
  three-dimensional fluid flows.
\newblock \emph{Physica D: Nonlinear Phenomena 149}, 4 (2001), 248--277.

\bibitem[HHFH16]{Haller:2016:Defining}
\textsc{Haller G., Hadjighasem A., Farazmand M., Huhn F.}:
\newblock Defining coherent vortices objectively from the vorticity.
\newblock \emph{Journal of Fluid Mechanics 795} (2016), 136--173.

\bibitem[HKR93]{ZD:Huttenlocher1993}
\textsc{Huttenlocher D.~P., Klanderman G.~A., Rucklidge W.~J.}:
\newblock Comparing images using the {Hausdorff} distance.
\newblock \emph{IEEE Transactions on Pattern Analysis and Machine Intelligence
  15}, 9 (1993), 850--863.

\bibitem[JBS08]{Janicke:2008:Brushing}
\textsc{Janicke H., Bottinger M., Scheuermann G.}:
\newblock Brushing of attribute clouds for the visualization of multivariate
  data.
\newblock \emph{Visualization and Computer Graphics, IEEE Transactions on 14},
  6 (nov.-dec. 2008), 1459 --1466.
\newblock \href {http://dx.doi.org/10.1109/TVCG.2008.116}
  {\path{doi:10.1109/TVCG.2008.116}}.

\bibitem[JBTS08]{Janicke:2008:Automatic}
\textsc{J{\"a}nicke H., B{\"o}ttinger M., Tricoche X., Scheuermann G.}:
\newblock Automatic detection and visualization of distinctive structures in 3d
  unsteady multi-fields.
\newblock \emph{Computer Graphics Forum 27}, 3 (2008), 767--774.

\bibitem[JWSK07]{Janicke:2007:Multifield}
\textsc{Janicke H., Wiebel A., Scheuermann G., Kollmann W.}:
\newblock Multifield visualization using local statistical complexity.
\newblock \emph{IEEE Transactions on Visualization and Computer Graphics 13}, 6
  (2007), 1384--1391.

\bibitem[KMM{\etalchar{*}}01]{Kniss:2001:Interactive}
\textsc{Kniss J., McCormick P., McPherson A., Ahrens J., Painter J., Keahey A.,
  Hansen C.}:
\newblock Interactive texture-based volume rendering for large data sets.
\newblock \emph{IEEE Comput. Graph. Appl. 21} (July 2001), 52--61.
\newblock URL: \url{http://dx.doi.org/10.1109/38.933524}, \href
  {http://dx.doi.org/http://dx.doi.org/10.1109/38.933524}
  {\path{doi:http://dx.doi.org/10.1109/38.933524}}.

\bibitem[MW95]{Martin:1995:High}
\textsc{Martin A.~R., Ward M.~O.}:
\newblock High dimensional brushing for interactive exploration of multivariate
  data.
\newblock In \emph{Proceedings of the 6th conference on Visualization '95}
  (Washington, DC, USA, 1995), VIS '95, IEEE Computer Society, pp.~271--.
\newblock URL: \url{http://portal.acm.org/citation.cfm?id=832271.833844}.

\bibitem[Pop04]{ZD:Popinet2004}
\textsc{Popinet S.}:
\newblock Free computational fluid dynamics.
\newblock \emph{ClusterWorld 2}, 6 (2004), 7.

\bibitem[QCX{\etalchar{*}}07]{Qu:2007:Visual}
\textsc{Qu H., Chan W.-Y., Xu A., Chung K.-L., Lau K.-H., Guo P.}:
\newblock Visual analysis of the air pollution problem in hong kong.
\newblock \emph{Visualization and Computer Graphics, IEEE Transactions on 13},
  6 (nov.-dec. 2007), 1408 --1415.
\newblock \href {http://dx.doi.org/10.1109/TVCG.2007.70523}
  {\path{doi:10.1109/TVCG.2007.70523}}.

\bibitem[SA92]{ZD:Spalart1992}
\textsc{Spalart P.~R., Allmaras S.~R.}:
\newblock A one equation turbulence model for aerodinamic flows.
\newblock \emph{AIAA Journal 94} (1992).

\bibitem[SSBW05]{Stockinger:2005:DEX:}
\textsc{Stockinger K., Shalf J., Bethel W., Wu K.}:
\newblock Dex: Increasing the capability of scientific data analysis pipelines
  by using efficient bitmap indices to accelerate scientific visualization.
\newblock In \emph{In International Conference on Scientific and Statistical
  Database Management (SSDBM} (2005), IEEE Computer Society Press, pp.~35--44.

\bibitem[STS06]{Sauber:2006:Multifield-Graphs:}
\textsc{Sauber N., Theisel H., Seidel H.-P.}:
\newblock Multifield-graphs: An approach to visualizing correlations in
  multifield scalar data.
\newblock \emph{IEEE Transactions on Visualization and Computer Graphics 12}
  (2006), 917--924.
\newblock \href
  {http://dx.doi.org/http://doi.ieeecomputersociety.org/10.1109/TVCG.2006.165}
  {\path{doi:http://doi.ieeecomputersociety.org/10.1109/TVCG.2006.165}}.

\bibitem[STW{\etalchar{*}}08]{Shi:2008:Visualizing}
\textsc{Shi K., Theisel H., Weinkauf T., Hege H.-C., Seidel H.-P.}:
\newblock Visualizing transport structures of time-dependent flow fields.
\newblock \emph{IEEE Computer Graphics \&amp; Applications} (2008).

\bibitem[TGK{\etalchar{*}}04]{ZD:Tricoche2004}
\textsc{Tricoche X., Garth C., Kindlmann G., Deines E., Scheuermann G., Ruetten
  M., Hansen C.}:
\newblock Visualization of intricate flow structures for vortex breakdown
  analysis.
\newblock In \emph{Proceedings of the Conference on Visualization'04} (2004),
  IEEE Computer Society, pp.~187--194.

\bibitem[W{\etalchar{*}}98]{ZD:Wilcox1998}
\textsc{Wilcox D.~C., et~al.}:
\newblock \emph{Turbulence Modeling for CFD}, vol.~2.
\newblock DCW industries La Canada, CA, 1998.

\bibitem[WPJH08]{Weldon:2008:Experimental}
\textsc{Weldon M., Peacock T., Jacobs G.~B., Helu M.}:
\newblock Experimental and numerical investigation of the kinematic theory of
  unsteady separation.
\newblock \emph{Journal of Fluid Mechanics 611} (September 2008), 1--11.

\bibitem[WYG{\etalchar{*}}11]{Wang:2011:Analyzing}
\textsc{Wang C., Yu H., Grout R., Ma K.-L., Chen J.}:
\newblock Analyzing information transfer in time-varying multivariate data.
\newblock In \emph{Pacific Visualization Symposium (PacificVis), 2011 IEEE}
  (march 2011), pp.~99 --106.
\newblock \href {http://dx.doi.org/10.1109/PACIFICVIS.2011.5742378}
  {\path{doi:10.1109/PACIFICVIS.2011.5742378}}.

\bibitem[WYM08]{Wang:2008:Importance-Driven}
\textsc{Wang C., Yu H., Ma K.-L.}:
\newblock Importance-driven time-varying data visualization.
\newblock \emph{Visualization and Computer Graphics, IEEE Transactions on 14},
  6 (nov.-dec. 2008), 1547 --1554.
\newblock \href {http://dx.doi.org/10.1109/TVCG.2008.140}
  {\path{doi:10.1109/TVCG.2008.140}}.

\end{thebibliography}

\end{document}